
\input phyzzx
\overfullrule 0pt

\PHYSREV

\def\refmark#1{$^{\scriptstyle#1}$}
 \doublespace
 \Pubnum{SLAC--PUB--6570\cr
  hep-ph-xxxxxxx \cr}
 \date{July, 1994}
 \pubtype{T/E}
 \titlepage
\vglue 1.5in
 \title{Exclusive Processes:
 Tests of Coherent QCD Phenomena
 and Nucleon Substructure at CEBAF \doeack}
 \author{Stanley J. Brodsky}
 \SLAC
 \bigskip
\vfill
\titlestyle{
Invited Overview Talk Presented at the \break
CEBAF Workshop on Electroproduction\break
Newport News, Virginia\break
April 14--16, 1994}
\vfill
\endpage

\def\vvec{\mathaccent"017E }

\def\kp{\vvec k_\perp{}}
\def\qp{\vvec q_\perp{}}
\def\lp{\vvec \ell_\perp{}}
\def\qps{\vvec q_\perp^{\,2}{}}

\def\M{{\cal M}}

\headline={\ifnum\pageno=1\firstheadline\else
\ifodd\pageno\rightheadline \else\leftheadline\fi\fi}
\def\firstheadline{\hfil}
\def\rightheadline{\hfil}
\def\leftheadline{\hfil}
        \footline={\ifnum\pageno=1\firstfootline\else\otherfootline\fi}
\def\firstfootline{\rm\hss\folio\hss}
\def\otherfootline{\hfil}
\font\tenbf=cmbx10
\font\tenrm=cmr10
\font\tenit=cmti10
\font\elevenbf=cmbx10 scaled\magstep 1
 1
 1

\line{\hfil }
\vglue 1cm
\hsize=6.0truein
\vsize=8.5truein
\parindent=3pc
\baselineskip=14pt
\centerline{\tenbf  EXCLUSIVE PROCESSES:}
\centerline{\tenbf  TESTS OF COHERENT QCD PHENOMENA}
\centerline{\tenbf
AND NUCLEON SUBSTRUCTURE AT CEBAF  }
\vglue 1.0cm
\centerline{\tenrm STANLEY J. BRODSKY}
\centerline{\tenit Stanford Linear Accelerator Center,
Stanford University}
\centerline{\tenit Stanford, CA 94309 USA}
\vglue 0.6cm
\centerline{\tenrm ABSTRACT}
\vglue 0.3cm
{\rightskip=3pc
 \leftskip=3pc
 \tenrm\baselineskip=12pt
 \noindent
Measurements of exclusive processes such as
electroproduction,
photoproduction, and Compton
scattering  are among the most
sensitive probes of proton structure and coherent
phenomena in
quantum chromodynamics.
The continuous electron beam at CEBAF, upgraded in
laboratory energy to 10--12 GeV, will allow a systematic study
of exclusive, semi-inclusive,
and inclusive
reactions in a kinematic range
well-tuned to the study of fundamental nucleon and nuclear substructure.
I also discuss the potential at CEBAF
for studying  novel QCD phenomena at the
charm production threshold, including
the possible production of nuclear-bound quarkonium.}
\vglue 0.6cm
\line{\elevenbf 1. Introduction:  Electroproduction at CEBAF \hfil}
\vglue 0.4cm

Quantum Chromodynamics has become
the central focus of particle and nuclear physics
since it potentially can describe all strong interactions in terms
of fundamental quark and gluon degrees of freedom.
Although there  have been
many empirical successes of QCD, crucial elements of the theory remain
unexplored.  For example, the fundamental structure of hadronic
wavefunctions, the nature of confinement,  and the structure of the
QCD vacuum
are understood only qualitatively at
best.
The study of hadronic wavefunctions has now become even more
critical in view of the need to understand exclusive and
inclusive charm and bottom hadronic
from first principles.  An important
challenge is to devise tests of the theory
and hadron structure at  the amplitude level which can
be made as systematically precise as possible.

\REF\MuellerQCD{%
See, \eg, the volume
{\it Perturbative Quantum Chromodynamics,}
        Edited by A.H. Mueller.
        World Scientific, 1989.}

A characteristic scale of quantum chromodynamics is the mean transverse
momentum of quarks within hadrons:
$ \VEV{k_T^2}^{1/2} \simeq
200 - 300 $ MeV.  Hadronic processes involving momentum transfer
much larger than this scale $Q^2 \gg  \VEV{k_T^2}^{1/2}$ can
be traced to underlying hard
scattering process $T_H(Q)$ involving the minimal number of quark and
gluons.  Because of asymptotic freedom, $T_H(Q)$ can be computed in
perturbation theory as an expansion in powers of $\alpha_s(Q^*)$, where
$Q^* = {\cal O}(Q).$ This observation is the basis of the QCD
factorization theorems, in which color confinement and the
non-perturbative effects of hadronic binding are isolated in terms of the
hadron structure functions and fragmentation functions in the case of
inclusive reactions, and distribution amplitudes $\phi_H(x,Q)$ in the
case of exclusive reactions.\refmark\MuellerQCD\
Processes which involve the hard scattering
of more than the minimum numbers of quarks or gluons are ``higher-twist";
\ie\ they are dynamically suppressed by powers of $\Lambda^2/Q^2.$

The extended  laboratory electron energy range
now being contemplated at CEBAF, up to 10-12 GeV, allows one to probe
QCD effects in the transition regime between coherent and incoherent
quark subprocesses.  In the case of inclusive electroproduction at CEBAF,
the dominant subprocess at large momentum transfer corresponds to
the electron scattering
on one of the quark constituents
of the target nucleon or nucleus.  Thus the deep inelastic cross
section $d\sigma (e N \to e' X)$ is to first approximation given by
the convolution of the hard scattering $d\sigma (e q \to e' q')$ cross
section, multiplied by target structure functions $G_{q/N}(x,Q),$ the
probability distributions describing the spin and flavor distributions of
quarks in the proton, neutron, or nuclei at light-cone momentum fraction
$x=k^+/ p^+=(k^0+k^3)/(p^0 + p^3)$.
On-shell kinematics allows
one to identify $x$ with the Bjorken
variable $x_{bj} = Q^2/2 p\cdot q.$
Since the beam is continuous  and the
number of produced hadrons is not enormous, one can measure the
complete final state in electroproduction at CEBAF,
and thus follow the evolution
of the produced quarks and
gluons from the hard subprocess and the remnant
quarks and gluons spectators from the target into final state hadronic
systems.  Since the momentum transfers is moderate,
one can also detect the non-factorizing effects of coherence,
such as the interference between subprocesses in which the
electron scatters on different quarks in the target.

At very high electroproduction energies, for example in electron-proton
collisions at HERA, deep inelastic
lepton scattering has only minimal sensitivity to
the valence parton structure of the proton.
{} From the
perspective of the proton rest frame, the high energy virtual photon
fluctuates into virtual $q \bar q$ system which then scatters on the
gluonic field of the target; \ie,
physics associated with
photon dissociation and central rapidity region processes.
Thus physics at HERA  focuses more on
the structure of the photon rather than resolving nucleon substructure.
In contrast, at CEBAF energies, the dominant electroproduction
physics is controlled
by the quark structure of the target nucleon or nucleus.

Monte-Carlo and string fragmentation programs are often used
to simulate the
main features of the final state
hadronization in electroproduction;
However the goal is to acquire
a fundamental understanding of hadronization
at the amplitude level. Studies at CEBAF have the potential
for studying the
basic physical processes involved in the processes in which a confined
quark or gluon turns into hadronic matter. In
the case of nuclear target, one can resolve the
effects of the background nuclear field such as quark energy loss
transverse
momentum smearing and co-mover interactions on the materialization
of the final state.

Although the energy range proposed for an upgraded CEBAF  is
well-tuned to resolving proton structure in QCD,
in general one also needs to take into account coherent effects and
multiparticle subprocesses.  For example, at moderate momentum transfers,
the electron will often interact with more than one target constituent;
\eg, $e q q \to e' q' q'.$ Since the two quarks can scatter coherently
as a bosonic system, they can produce a large longitudinal cross section
$R= \sigma_L/\sigma_T.$ The two quarks together carry a large
fraction of the target momentum, and thus such higher twist contributions
can actually dominate the electroproduction cross section at large $x
\sim 1.$

\REF\BBKM{%
A. Brandenburg, S. J. Brodsky, V. V. Khoze, and D. M\"uller,
SLAC-PUB-6464 (1994), to be published
in Phys. Rev. Lett.}

Although higher twist terms may complicate the physical interpretation of
electroproduction at CEBAF, they are important and interesting topics in
their own right.
For example, Brandenburg, Khoze, M\"uller, and I\refmark\BBKM~ have
recently shown that measurements of the $\cos \phi$ and $\cos 2 \phi$
azimuthal angular dependence of the lepton in the Drell-Yan process $H N
\to \ell \bar \ell X$ with $H = \pi, K, \bar p$ and $p$ at forward $x_F$
are sensitive
to the shape of the projectile's distribution
amplitudes.  One can analytically
cross these predictions to obtain a theory of meson
electroproduction including higher twist contributions where the meson
interacts directly within the hard subprocess.  In the case of meson
electroproduction, $\phi$ is the azimuthal angle between the lepton
scattering plane and the meson production plane.  Thus
measurements at CEBAF
of the full azimuthal and polar angular distribution of the
lepton system in meson electroproduction and lepton pair hadroproduction
should
provide an important measure of the structure of hadrons at the amplitude
level.

\REF\BL{For a review of the theory of exclusive processes
in QCD and additional references see
S. J. Brodsky and G. P. Lepage in {\it Perturbative
Quantum Chromodynamics}, edited by A. Mueller
(World Scientific, Singapore, 1989).}

\REF\BF{%
S. J. Brodsky and G. R. Farrar,
{\it  Phys. Rev.} {\bf D11} (1975) 1309.}

\REF\CT{%
S. J. Brodsky and A. H. Mueller,
{\it Phys. Lett.} {\bf 206B} (1988) 685, and references therein;
G. Bertsch, S. J. Brodsky,
A. S. Goldhaber, J.F. Gunion,
{\it Phys. Rev. Lett.} {\bf 47} (1981) 297.}

The study of exclusive reactions at CEBAF such as elastic electron-proton
scattering, real and virtual Compton scattering, and meson
electroproduction, provides a complimentary measure of nucleon structure
to the purely inclusive studies.\refmark\BL~
An analogy is an electron microscope,
where information from both elastic and inelastic scattering are combined
to generate the image of the target.  In exclusive reactions in QCD all
of the constituents of the scattered hadrons must be rearranged from the
initial to final state.
Dimensional counting rules\refmark\BF~
show that the leading
subprocesses to order $ Q^{-1}$ involve the minimum number of incident
and final constituents.
Furthermore, the valence quarks
exchange their hard momenta when their transverse separations are small:
$b^i_\perp = {\cal O} (Q^{-1}).$
Thus large momentum transfer exclusive reactions
are controlled by the hadron distribution amplitudes $\phi_H(x_i,Q),$
the valence light-cone Fock
wavefunction at small transverse separation.
This leads  to the remarkable ``color transparency''\refmark\CT~
property of QCD, since
such small color singlet fluctuations have only minimal initial
and final state interactions as they transit through nuclei.
In addition, the study of exclusive pion and kaon electroproduction
over a large kinelatic range of energy and
momentum transfer is necessary in order to reliably determine
the spacelike meson form factors.
A more detailed discussion will be presented in Section 4.

\REF\Fang{%
E665 Collaboration (G. Y. Fang, \etal),
FERMILAB-CONF-94-041-E, (1994).
Presented at the {\it 23rd International
Multiparticle Dynamics Symposium},
1993, Aspen.}

\REF\BFGMS{%
S. J. Brodsky, L. Frankfurt, J. F.
Gunion, A. H. Mueller, and M. Strikman,
SLAC-PUB-6412, (1994). (To be published in {\it Phys. Rev. D.})}

In the case of
electroproduction at CEBAF one can use a nuclear target as a ``color
filter" to separate large
and small structure events.  Recently, the E665
group at Fermilab\refmark\Fang~
has reported preliminary results on coherent and incoherent $\rho$
leptoproduction in nuclei.  At low momentum transfer, the $\rho$ vis
strongly absorbed, as in conventional Glauber theory; however, as $Q^2$
increases beyond a few GeV$^2$ the reactions tend to occur uniformly
throughout the nuclear volume, as predicted by color transparency.  Thus
diffractive muo-production of $\rho$ mesons occurs in a nuclear target
without final absorption of the $\rho$ in the nucleus.  More generally,
by using nuclear targets, one can change the hadronic environment and
study not only the shadowing and antishadowing of nuclear structure
functions, but also the influence of the nuclear field on evolution of
the final state hadronic system, including the induced radiation of the
outgoing quarks.\refmark\BFGMS  It will be clearly interesting to
trace the color transparency effects seen in vector meson
leptoproduction at Fermilab to the lower
CEBAF energy range where the formation times are moderate.

\REF\Ar{%
R. L. Anderson \etal, {\it Phys. Rev. Lett.} {\bf 30} (1973) 627.}

\Picture\photo\width=\hsize
\caption{\narrower\singlespace
Comparison of photoproduction data with the dimensional counting
power-law prediction. The data are summarized in Ref. \Ar.
}
\savepicture\photopic
\midinsert
\vskip 5in
\photopic\endinsert

A central prediction of perturbative QCD for
exclusive electroproduction at large momentum transfer
is fixed center-of-mass angle scaling:
$${d\sigma\over dt}{(\gamma^* p \to M B)} = {f(t/s,Q^2/s) \over s^N}.$$
The nominal scaling power  $N \simeq 7$ follows from dimensional
counting: there are 4 incident and 5 outgoing elementary fields.
One important test of this scaling is shown in Fig. \photo\
for pion photoproduction $\gamma p \to
\pi^+ n$  at $\theta_{cm} = \pi/2.$
The nominal $s^{-7}$ predicted power law
behavior is consistent with experiment over the energy range
contemplated at CEBAF.  This scaling behavior needs to
be checked systematically in electroproduction, for example,
as a function of virtual photon mass and
polarization, and the angular and energy range.
The leading power should correspond to helicity amplitudes
which conserve the total
hadron helicity from the initial and final state,
independent of the photon polarization.
Hadron helicity conservation is discussed in more detail in Section 4.

\REF\Carroll{%
A. Carroll, Presented at the {\it Workshop
on Exclusive Processes at High Momentum Transfer},
Elba, Italy,  1993; C. White \etal, BNL-49059, (1993);
B. R. Baller \etal, {\it Phys. Rev. Lett.} {\bf 60} (1988) 1118.}

\REF\BBG{%
J. F. Gunion, S. J. Brodsky, and R. Blankenbecler,
{\it Phys. Rev.} {\bf D8} (1973) 287.}

In general PQCD dimensional counting  has been shown
to be good guide to the scaling behavior of general
fixed angle two-body scattering reactions.\refmark\BL~
A systematic study of meson-baryon reactions has recently been
completed at Brookhaven.\refmark\Carroll\
The large relative normalization of
large angle cross sections such as $K^+ p \to K^+ p$  compared
to $K^- p \to K^- p$  shows that the dominant interaction controlling
exclusive processes at large momentum transfer involves the
interchange of the valence quarks\refmark\BBG~
ra\-ther than multiple gluon
exchange.   Other important tests involve
exclusive two-photon reactions such as $\gamma^* \gamma \to M^0$,
$\gamma \gamma \to M \bar M$, and proton-proton annihilation.
The general success of dimensional counting in the
fixed angle domain is evidence that leading twist
PQCD mechanisms dominate exclusive amplitudes
and form factors at momentum transfers $Q^2 \sim 5$ GeV$^2.$

\REF\LS{%
J. Botts and G. Sterman,
{\it Nucl. Phys.} {\bf B325} (1989) 62;
{\it Phys. Lett.} {\bf B224} (1989) 201;
J. Botts, J.-W. Qiu, and G. Sterman,
{\it Nucl. Phys.} {\bf A527} (1991) 577.
H.~N.~Li and G. Sterman,
{\it Nucl. Phys.} {\bf  B381} (1992) 129.
H. N. Li, Stony Brook preprint ITP-SB-92-25 (1991).
}

\REF\SS{%
M. G. Sotiropoulos and  G. Sterman
ITP-SB-93-59 (1993),
ITP-SB-93-83 (1994).}

The structure of hadron wavefunctions in terms of their quark and gluon
degrees of freedom at the amplitude level remains one of the most
important frontiers in QCD studies.  The natural formulation of hadron
wavefunctions is the light-cone Fock expansion.\refmark\BL~
As noted above, the basic quantity which characterizes the part
of the hadron bound state which enters exclusive hard-scattering
subprocesses is the gauge and frame-independent distribution
amplitude $\phi(x_i,Q)$ which in turn describes the valence quark
structure of the hadrons at impact separation $b^i_\perp = {\cal O}
(Q^{-1})$ is a function of the light-cone momentum fractions.  The work
of Sterman \etal,\refmark\LS~  has  shown in detail
how Sudakov suppression of large size configurations of the hadron
wavefunctions are suppressed in large momentum transfer exclusive
processes, confirming the validity of the PQCD description of these
processes and the corresponding predictions of QCD color transparency.
The interrelation of the Landshoff triple-gluon contributions to elastic
proton-proton scattering to hard scattering PQCD mechanisms
such as quark interchange has now been
clarified by Sterman and Sotiropoulos.\refmark\SS

\vglue .13 in
\vbox{
QCD sum rule methods and lattice gauge theory now supply important
theoretical constraints on the form of the distribution amplitudes,
although the reliability of these predictions is unknown.  There is now
much theoretical work exploring other non-perturbative QCD methods such
as light-cone Hamiltonian diagonalization.  One can also derive
constraints on hadron light-cone wavefunctions from their static
properties such as the baryon magnetic moments and their axial couplings.
}
 \goodbreak

\vglue 0.6cm
\line{\elevenbf 2. The Charm Threshold in Electroproduction \hfil}
\vglue 0.4cm

One of the most interesting physics areas which can be studied at CEBAF
at electron beam energies above 8 GeV will be the onset of charm
electroproduction.  The threshold virtual photon energy for the lowest
hidden charm system $\gamma^* p \to \eta_c(2.9788 {\rm GeV}) p(0.9383
{\rm
GeV})$ is $\nu_{th} = q \cdot p /M = 7.707$ GeV $+ (Q^2/2 M_p).$
The production of open charm $\gamma^* p \to D^0 (1.8645 {\rm GeV})
\Lambda_c (2.2849 {\rm GeV})$ begins at $\nu_{th} = 8.7057$ GeV $
+ (Q^2/ 2 M_p).$
In the threshold regime one probes extreme configurations
of the proton target quark structure as it strains to produce the new
heavy systems.  The study of this physics also has implications for the
charm structure function at large $x_{bj}$ and the threshold production
of heavy systems such as beauty, top, and supersymmetric particles.

\REF\LMS{%
M. Luke, A. V. Manohar, M. J. Savage,
{\it Phys. Lett.} {\bf B288} (1992) 355.}

\REF\BDeTS{%
S. J. Brodsky, and G. F. de Teramond, and I. A. Schmidt,
{\it Phys. Rev. Lett.} {\bf 64} (1990) 1011.}

\REF\Wilkin{%
See, \eg,  C. Wilkin,
{\it Phys. Rev.} {\bf C47} (1993) 938.}

Although the charm electroproduction cross section is inevitably
suppressed at threshold by phase space, there is reason to believe that
the production rate will be substantially enhanced by dynamical effects
in non-perturbative QCD.  Because of the disparate charmonium and proton
size scales, one can classify and compute their two-gluon exchange
couplings using the operator product expansion.
Luke, Manohar, and Savage\refmark\LMS~
have shown that the scalar part of the two-gluon exchange interaction
is related to the trace
of the energy momentum tensor and thus its coupling to
nucleons or nuclei is proportional to the target mass.
The two-gluon coupling to the small size charmonium
state can be computed using conventional potential models.
This analysis leads to the remarkable prediction that there is
a strong QCD van der Waal attraction
of the quarkonium state to ordinary hadrons at small
relative velocity.  DeTeramond, Schmidt, and I\refmark\BDeTS~
have argued that the
QCD van der Waal effects at low relative velocity could be
sufficiently strong as
to bind charmonium states to ordinary hadrons or light nuclei.
Such nuclear-bound quarkonium states could show up as narrow $s-$channel
resonance in electroproduction: $\gamma^* p \to (\eta_c p),$ $\gamma^* d
\to (\eta_c d),$ $\gamma^* p \to (J/\psi p),$ etc.  just below the
charmonium production threshold.
This could be a very interesting experiment CEBAF.
However, it should be emphasized that even if the QCD van
der Waals force is not sufficient to actually form bound states, one
still expects to see strong threshold effects in the production cross
section perhaps similar to the enhancements that have been observed
for $\eta$ production at threshold.\refmark\Wilkin\

Note that even though the rate at
threshold may be small, the cross section
can be enhanced by using nuclear Fermi-motion
to effectively increase the available
energy.  For example, the anti-deuteron was first observed below
the nominal threshold energy in the 1960s at Brookhaven by Ting and
Lederman using heavy nuclei as targets.  It is also interesting
to use the charm
threshold to measure the extreme limits of the Fermi momentum spectrum,
since its origin  involves nuclear short-range interactions.

\Picture\figinter\width=\hsize
\caption{\narrower\singlespace
Quark interchange contribution to charm electroproduction.
}
\savepicture\figinterpic
\midinsert
\vskip 2in
\figinterpic\endinsert

In the case of open charm, the simplest electroproduction mechanism is
quark interchange, as illustrated in Fig. \figinter.
The interchange amplitude can be written in an elegant form as
a convolution over valence light-cone
wavefunctions: $$\int {d^2k_\perp dx \over 16 \pi^3} \psi^\dagger_D
(x,k_T+ x r_\perp) \psi^\dagger_{\Lambda_c}(x,k_T+(1-x)q_\perp) \Delta
\psi_{\gamma^*}(x,k_T+ x r_\perp + (1-x)q_\perp) \psi_p(x,k_T),$$ where
$q^2_\perp = -t, r^2_\perp = -u$ and
$\Delta$ is the inverse of the light-cone energy denominator.  The
complete analysis is given in Ref. \BBG.

There are a number of experiments which indicate that non-perturbative
QCD mechanisms are necessary for understanding heavy quark production in
the regimes where either the charm system is produced at extreme
kinematic configurations such as large $x_F$ or large $x_{bj}$ or at
small relative velocity to other quarks:
\REF\charmrev{%
For a review of intrinsic heavy quark phenomena
and further references, see S.~J.~Brodsky, SLAC-PUB-6304,
{\it CCAST Symposium on Particle Physics at the Fermi
Scale}, Beijing, China,  (1993); and
R. Vogt, S. J. Brodsky, and P. Hoyer
{\sl Nucl. Phys.} {\bf B383} (1992) 643.}
\pointbegin The anomalously high $c(x)$ distribution measured at large
$x_{bj}$ by EMC.\refmark\charmrev~
The CERN measurements disagree with photon-gluon fusion
by a factor of 20 to 30 at $Q^2 = 75$ GeV$^2$ and $x_{bj} = 75$ GeV$^2$.
\point In the case of $J/\psi$ hadroproduction from pion beams, the CERN
experiment NA-3 has reported a strong excess of quarkonium at large $x_F$
with a non-factorizing nuclear dependence.  In addition, the Fermilab
Chicago-Iowa Princeton group has reported an anomalously sudden change in
polarization of the $J/\psi$ at large $x_F$ in $\pi N \to \mu^+ \mu^- X$.
The dramatic shift to longitudinal polarization is inconsistent with
leading order QCD predictions.
\point An interesting unresolved issue is the leading particle effect in
charmed hadron production.  The quark structure of leading $D$ mesons has
been shown to depend strongly on the valence quantum numbers of the beam
hadron in direct conflict with the factorization principle at the heart
of most perturbative QCD predictions.  The mechanisms in which the beam
quarks and heavy quarks coalesce is at the heart of hadronization
dynamics, and much more critical work will be needed especially in the
production of b-quark systems.
\point There are other signals for anomalous charm baryon hadroproduction
at large $ x_F $, including the reports of $\Lambda_c$ production from
E-400 at Fermilab using neutron beams and the measurements of WA-62 from
CERN which observed charm-strange baryons using hyperon beams.  There are
also measurements from NA-3 at CERN which show that double $J/\psi$ pairs
are hadro-produced only at large $x_F.$
\point The anomalously strong nuclear dependence of large $x_F$ $J/\psi$
hadroproduction, as reported by NA10 and E789 are in direct contradiction
to leading-twist PQCD factorization.

\REF\BHMT{%
S. J. Brodsky, P. Hoyer, A. H. Mueller, W-K. Tang,
{\it Nucl. Phys.} {\bf B369} (1992) 519.}

Much of the above physics can be accounted for by the picture of Hoyer,
Mueller, Tang and myself,\refmark\BHMT~
where the hadronization in a high energy
collision occurs in the following novel way: the heavy quark system is
first formed as a virtual fluctuation as a light-cone Fock state
component in the incoming hadron wavefunction; a light spectator quark is
then stripped away in the target leaving the $Q \bar Q$ system to
hadronize into the final heavy hadrons.  This type of intrinsic heavy
quark picture also explains the excess of charm quarks seen in the EMC
measurements of the charm structure function of the nucleon.  This new
picture of hadron formation opens up a whole new avenue for studying the
far-off-shell structure of hadrons.  It is thus critical that a new
measurement of the charm and beauty structure functions be performed.

Measurements of charm  electroproduction near threshold at CEBAF
should provide new insights into the collective multi-quark
mechanisms needed to understand the charm production anomalies.

\vglue 0.6cm
\line{\elevenbf 3.  Virtual Compton Scattering\hfil}
\vglue 0.4cm

The Compton scattering process $\gamma p \to \gamma p$ is the fundamental
way to ``look'' at proton structure.  Virtual Compton scattering
$\gamma^* p \to \gamma p$ is
particularly interesting to measure
at CEBAF since it can
be probed as a function of the photon's transverse or longitudinal
polarization, the target polarization, over a large domain of kinematics
$s, t, u,$ and photon virtuality $Q^2= -q^2.$

\REF\BCG{%
S.  J.  Brodsky, F.  E.  Close, J.  F.  Gunion, {\it Phys.
Rev.}  {\bf D5} (1972) 1384.}

\REF\BGJ{%
S.  J.  Brodsky, J.  F.  Gunion, R.  Jaffe (SLAC), {\it Phys.
Rev.} {\bf D6} (1972) 2487.}

It should be noted that the cross section for the process $e p \to e' p
\gamma$ receives contributions not only from virtual Compton scattering,
but also from Bethe-Heitler bremmstrahlung from the scattered electron.
The two processes lead to the same final state, and thus they interfere.
The Bethe-Heitler process is completely determined from elastic $e p$
scattering and is purely real.  Thus one can use the interference between
the Compton and bremmstrahlung processes to determine the real part of
the Compton amplitude.\refmark\BCG\ In the case of deep inelastic Compton
scattering $e p \to e \gamma X$, one can use the same interference
effect to deduce new structure functions and sum rules proportional to
the sum of quark charges cubed.\refmark\BGJ\

\REF\BHP{%
S.  J.  Brodsky, A.  C.  Hearn, R.  G.  Parsons, {\it Phys.
Rev.} {\bf 187} (1969) 1899.}

\REF\DamGil{%
M. Damashek, F. J. Gilman, {\it Phys.  Rev.} {\bf D1} (1970) 1319.}

There are many different physics aspects of virtual Compton scattering
depending on the accessed kinematical domain.
\pointbegin
In the case of low energy virtual Compton scattering with $s = (q + p)^2
\simeq M^2_{N^*}$ one can study the s-channel effects of baryon
resonances in the Compton amplitude and their relative coupling as a
function of photon virtuality.\refmark\BHP
\point
In the Regge limit $s \gg -t$ and fixed $Q^2$ one can use the Regge pole
analysis, as in the paper of Damashek and Gilman.\refmark\DamGil~ Each
Compton helicity amplitude has the form of a sum over $t-$channel Regge
exchange contributions: ${\cal M} = \Sigma_R s^{\alpha_R(t)}
\beta(t,Q^2). $ In Compton scattering one can have contribution from all
$C$-even exchanges: the diffractive Pomeron contributions,the pion and
$A_2$ and $f^0$ $C=+$ Reggeon trajectories.  In addition, QCD predicts a
special contribution which cannot occur in hadron-hadron scattering: a
$j=0$ fixed pole, the Kronecker $\delta_{j,0}$ contribution, which can be
traced to the presence of quark Z-graphs.
\point
An important feature of the Regge theory which is
testable in virtual Compton scattering is that the Regge
trajectory $\alpha_R(t)$ must be independent of $Q^2$ at fixed $-t.$ Only
the residue $\beta(t,Q^2)$ can depend on the photon virtuality.  In fact,
one expects that all of the normal trajectories have decreased couplings
to the virtual; Compton amplitude as $Q^2$ increases, leaving the $j=0$
fixed pole as the dominant and surviving contribution to the amplitude.
This special contribution to Compton scattering gives an energy
independent contribution to the real part of the $\gamma^* p \to \gamma
p$ amplitude.\refmark\BCG\  The $t-$dependence of the $j=0$ fixed pole
amplitude is expected to be similar to that of the helicity-conserving
Dirac form factor of the proton.
\REF\BTT{%
S.  J.  Brodsky, W.-K.  Tang, C.  B.  Thorn, {\it Phys.  Lett.}
{\bf B318} (1993) 203.}
\point
As the momentum transfer squared to the proton increases $-t,$ the
Pomeron trajectory is expected to stay at $\alpha_p(-t) \simeq 1.$ The
non-singlet Regge trajectories are predicted to decrease monotonically to
$\alpha_R(-t) \simeq 0.$ For a recent discussion and further references
see Ref. \BTT.
\point
In the large momentum transfer domain with fixed $\cos \theta_{cm}$ the
virtual exclusive Compton amplitude $\gamma^* p \to \gamma p $ can be
analysed using perturbative QCD factorization.  Detailed QCD predictions
have been made by Kronfeld and Nizic, Hyer, and Gunion \etal.  This will
be discussed in detail in Section 5.  In addition at
CEBAF, one may be able to test
these predictions as a function of photon polarization and virtuality.

\REF\sjbelba{%
Part of this section was also presented at the at the {\it Workshop
on Exclusive Processes at High Momentum Transfer},
Elba, Italy,  1993.
}

\REF\refI{%
S.~D.~Drell and T.~M.~Yan, {\it Phys. Rev. Lett.} {\bf 24} (1970) 181.
}

\REF\DLCQ{%
S. J. Brodsky and H. C. Pauli
in  {\it Recent Aspects of Quantum Fields},
H.~Mitter and H.~Gausterer, Eds.;
Lecture Notes in Physics, Vol. 396,
Springer-Verlag, Berlin, Heidelberg, (1991), and reference therein.}

\REF\refD{%
G. P. Lepage and S. J. Brodsky, {\it Phys. Rev.} {\bf D22},
2157 (1980); {\it Phys. Lett.} {\bf 87B} (1979) 359;
{\it Phys.~Rev.~Lett.} {\bf 43} (1979) 545, 1625E.
}

\REF\refP{%
General QCD analyses of exclusive processes are given in Ref. \refD,
S.~J.~Brodsky and G.~P.~Lepage, SLAC-PUB-2294,
presented at the {\it Workshop on Current Topics in High Energy Physics},
Caltech (Feb.~1979), S.~J.~Brodsky, in the {\it Proceedings
of the La Jolla
Institute Summer Workshop on QCD}, La Jolla (1978),
A.~V.~Efremov and A.~V.~Radyushkin, {\it Phys.~Lett.}
{\bf B94} (1980) 245,
V.~L. Chernyak, V.~G.~Serbo, and A.~R.~Zhitnitskii,
{\it Yad.~Fiz.} {\bf 31,} (1980) 1069,
S.~J.~Brodsky, Y.~Frishman, G.~P.~Lepage, and C.~Sachrajda,
{\it Phys. Lett.} {\bf 91B} (1980) 239, and
A.~Duncan and A.~H.~Mueller, {\it Phys.~Rev.} {\bf D21} (1980) 1636.
}

\REF\refQ{%
QCD predictions for the pion form factor at asymptotic $Q^2$ have
ben given by
V.~L.~Chernyak, A.~R.~Zhitnitskii, and V.~G.~Serbo,
{\it JETP Lett.} {\bf 26}
(1977) 594,   D.~R.~Jackson, Ph.D. Thesis, Cal Tech (1977),
and G.~Farrar and D.~Jackson,
{\it Phys.~Rev.~Lett.} {\bf 43} (1979) 246; and Ref. \refP.
See also A.~M.~Polyakov,
{\it Proceedings of
the International Symposium
on Lepton and Photon Interactions at High Energies},
Stanford (1975), and G.~Parisi, {\it Phys.~Lett.} {\bf 84B} (1979) 225.
See also
S. J. Brodsky and G. P. Lepage, in {\it High Energy Physics--1980},
{\it Proceedings of the XXth International Conference}, Madison,
Wisconsin, edited by L. Durand and L. G. Pondrom (AIP, New York,
1981); p. 568.
A. V. Efremov and A. V. Radyushkin, Rev. Nuovo Cimento {\bf 3},
1 (1980); and Ref. \refP.
V. L. Chernyak and A. R. Zhitnitskii,
{\it JETP Lett.} {\bf 25} (1977) 11;
M. K. Chase, {\it Nucl. Phys.} {\bf B167} (1980) 125.
}

\REF\CZ{%
V. L. Chernyak and A. R. Zhitnitskii, {\it
Phys. Rept.} {\bf 112} (1984) 173;
V. L. Chernyak, A. A. Oglobin, and I. R. Zhitnitskii,
{\it Sov. J. Nucl. Phys.} {\bf 48} (1988) 536;
I. D. King and C. T. Sachrajda,
{\it Nucl. Phys.} {\bf B297} (1987) 785;
M. Gari and N. G. Stefanis, {\it Phys. Rev.} {\bf D35} (1987) 1074;
and references therein.
}

\REF\refX{%
S.~J.~Brodsky, Y.~Frishman, G.~P.~Lepage and C.~Sachrajda,
Ref. \refP.  M. E. Peskin, {\it Phys. Lett.} {\bf 88B}  (1979) 128.
}

\REF\Miller{See, for example,
B. K. Jennings and  G.A. Miller
DOE-ER-40427-00-N93-11, (1993) and
{\it Phys. Lett.}{\bf  B236} (1990) 209;
G. R. Farrar, H. Liu,  L. L. Frankfurt, and M. I.
Strikman, {\it Phys. Rev. Lett.} {\bf 61} (1988) 686;
N. N. Nikolaev and B. G. Zakharov,
{\it Z. Phys. }{\bf C49} (1991) 607;
L. L. Frankfurt, M. I. Strikman, and M. B. Zhalov,
preprint (1993);
J. P. Ralston and B. Pire, {\it Phys. Rev. Lett.} {\bf 61} (1988) 1823,
and in the {\it Proceedings  of the 1989
24th  Rencontre de Moriond} (1989).}

\REF\BBL{%
S. J. Brodsky and G. P. Lepage,
{\it Phys. Rev.} {\bf D24} (1981) 1808.}

\REF\Nizic{%
B. Nizic, {\it Fizika} {\bf  18} (1986) 113.}

\REF\BREV{%
For a  review of exclusive two-photon processes, see
S. J. Brodsky,
{\it Proceedings of the  Tau-Charm
Workshop}, Stanford, CA (1989).
}

\REF\FARRAR{%
 G. R. Farrar, {\it et al}. {\it Nucl. Phys.}
{\bf B311} (1989) 585. }

\REF\GUNION{%
D. Millers and J. F. Gunion,
{\it Phys. Rev.} {\bf D34} (1986) 2657.}

\REF\KN{%
A. N. Kronfeld and B. Nizic, {\it Phys. Rev.} {\bf D44}
(1991) 3445;  B. Nizic, {\it Phys. Rev.} {\bf D35}(1987) 80.}

\REF\HYER{%
T. Hyer, {\it Phys. Rev.} {\bf  D47} (1993) 3875.}

\REF\INT{%
S. J. Brodsky, F. E. Close, J. F. Gunion,
{\it Phys. Rev.} {\bf D6} (1972) 177.}

\REF\Shupe{%
M. A. Shupe, {\it et al.},
{\it Phys. Rev.} {\bf D19} (1979) 1921.}

\REF\Stefanis{%
M. Bergmann and N. G. Stefanis, Bochum preprints
RUB-TPH-36/93, RUB-TPH-46/93, and RUB-TPH-47/93.}

\REF\SJBexcl{%
S. J. Brodsky, in the {\it Proceedings of the
Topical Conference on Electronuclear Physics with Internal
Targets}, Stanford, (1989).}

\REF\NNN{%
B. Z. Kopeliovich, J. Nemchick, N. N. Nikolaev, B. G. Zakharov,
{\it Phys. Lett.} {\bf B309} (1993) 179.}

\REF\Hel{%
G. P. Lepage and S. J. Brodsky, {\it Phys. Rev.} {\bf D24} (1981) 2848.}

\REF\Stoler{%
P. Stoler, {\it Phys. Rev.} {\bf D44} (1991) 73, {\it Phys. Rev. Lett.}
{\bf 66} (1991) 1003.}

\REF\ILS{%
N. Isgur and C. H. Llewellyn Smith,
{\it Phys. Rev. Lett.} {\bf 52} (1984) 1080;
{\it Phys. Lett} {\bf B217} (1989) 535.}

\REF\RAD{%
A. V. Radyushkin, {\it Nucl. Phys.} {\bf  A532} (1991) 141.}

\REF\DM{%
A. Duncan, and A. H. Mueller, {\it Phys. Lett.} {\bf 90B} (1980) 159.}

\REF\BG{%
S. J. Brodsky and J. F. Gunion,
{\it Phys. Rev. Lett.} {\bf 37} (1976) 402.}

\REF\SM{%
A. Szczepaniak and  L. Mankiewicz,
{\it Phys. Lett.} {\bf B266} (1991) 153.}

\REF\DMueller{%
D. M\"uller, SLAC-PUB (1993).}

\REF\Nagoya{%
S. J. Brodsky, I. A. Schmidt,
{\it Phys. Lett.} {\bf  B234} (1990) 144, and references therein;
S. J. Brodsky, in the {\it Proceedings of the
 International Symposium  on High-Energy
Spin Physics,} Nagoya, Japan,
(1992).}

\REF\HIGHER{%
S. J. Brodsky, E. L. Berger,  G. Peter Lepage,
{\it Proceedings of the Drell-Yan Workshop},
Fermilab (1982); E. L. Berger and S. J. Brodsky, {\it Phys. Rev. Lett.}
{\bf 42}  (1979) 940. For a recent analysis and additional
references see S. S. Agaev,
{\it  Z. Phys.} { \bf C57} (1993) 403.}

\REF\conway{%
See, \eg, J. S. Conway \etal, {\it Phys. Rev.} {\bf D39}  (1989) 92.}

\REF\BGMW{%
C. Greub, D. Wyler, S. J. Brodsky, and C. T. Munger,
SLAC-PUB-6487, (1994).}

\REF\blj{%
S.J. Brodsky, C.-R. Ji and G.P. Lepage,
{\it Phys. Rev. Lett.} {\bf 51} (1983) 83.
}

\REF\BC{%
S. J. Brodsky, B. T. Chertok,
{\it Phys. Rev.} {\bf  D14} (1976)  3003.}

\REF\Aan{%
S.~J.~Brodsky and J.~R.~Hiller,
{\it Phys.~Rev.} {\bf C28} (1983) 475.
}

\REF\Aam{%
C.~R.~Ji and S.~J.~Brodsky,
{\it Phys. Rev.} {\bf D34} (1986) 1460;
{\bf D33} (1986) 1951, 1406, 2653.
For a review of multi-quark evolution, see
S.~J. Brodsky, C.-R.~Ji, SLAC-PUB-3747,
(1985).
}

\REF\Holt{%
J.~Napolitano \etal,  ANL preprint PHY--5265--ME--88 (1988). }

\REF\Lee{%
T.~S.-H.~Lee, ANL preprint (1988).
}%

\REF\Myers{%
H.~Myers \etal, {\it Phys.~Rev.} {\bf 121} (1961) 630;
R.~Ching and C.~Schaerf, {\it Phys. Rev.} {\bf 141} (1966) 1320;
P.~Dougan \etal, {\it Z. Phys. A} {\bf  276} (1976) 55.}

\REF\KOD{%
L.A. Kondratyuk and M. G. Sapozhnikov, Dubna preprint E4-88-808.}

\REF\Krisch{%
For a summary of the spin correlation data see
A. D. Krisch,
{\it  Nucl. Phys. B (Proc. Suppl.)} {\bf  25B} (1992) 285.}

\REF\pinch{%
See, for example,  J. P. Ralston and B. Pire,
{\it Phys. Rev. Lett.} {\bf  49} (1982) 1605;
C. E. Carlson, M. Chachkhunashvili,
F. Myhrer, {\it Phys. Rev.} {\bf  D46} (1992) 2891;
G. P. Ramsey,   D. Sivers,
{\it Phys. Rev.} {\bf D47} (1992) 93; and references therein.}

\REF\Lands{%
P. V. Landshoff, {\it Phys. Rev} {\bf D10} (1974) 1024.}

\REF\Lip{%
S. J. Brodsky, C. E. Carlson, H. J. Lipkin,
{\it Phys. Rev.} {\bf D20} (1979) 2278.}

\REF\Sivers{%
Presented at the INT -
Fermilab Workshop on {\it Perspectives of High Energy Strong Interaction
Physics at Hadron Facilities} (1993).}

\REF\BDeT{%
S. J. Brodsky and G. F. de Teramond,
{\it Phys. Rev. Lett.} {\bf 60} (1988) 1924.}

\REF\Hepp{%
See S. Heppelmann,
{\it Nucl. Phys. B, Proc. Suppl.} {\bf  12} (1990) 159, and
references therein.}

\REF\NE{A. Lung, Presented at the {\it Workshop
on Exclusive Processes at High Momentum Transfer},
Elba, Italy,  1993.}

\REF\Tuan{%
S. J. Brodsky,  G. Peter Lepage, S. F.Tuan,
{\it Phys. Rev. Lett.} {\bf  59} (1987) 621, and references therein.}

\REF\Kirschner{%
R. Kirshner and L. N. Lipatov, {\it Sov. Phys.
JETP} {\bf 56} (1982) 266; {\it Nucl. Phys. }{\bf B213} (1983) 122.}

\REF\Savit{%
R. Blankenbecler, S. J. Brodsky, J. F. Gunion, and
R. Savit, {\it Phys. Rev.} {\bf  D8} (1973) 4117.}

\REF\Thorn{%
S. J. Brodsky, W-K. Tang, and C. B. Thorn, SLAC-PUB-6227 (1993).}

\vglue 0.6cm
\line{{\elevenbf 4.
Exclusive Processes and the Structure of Hadrons}\refmark\sjbelba\
\hfil}
\vglue 0.4cm

The analysis of exclusive hadronic amplitudes such as  form factors,
electroweak transition matrix elements, and
two-body scattering amplitudes has remained  among
the most challenging computational problems
in quantum chromodynamics. The physics of exclusive amplitudes
clearly depends
on the fundamental relativistic structure of the hadrons
as well as the dynamics governing quark and gluon propagation,
QCD vacuum structure, Regge behavior, and color confinement.
Numerical predictions
for  exclusive processes involving low momentum transfer
are beginning to be obtained
from lattice gauge theory and QCD sum rules.
However, the most interesting insights into hadron structure at the
amplitude level and the most
transparent connections to the underlying QCD physics emerges
at high momentum transfer where  perturbative analyses for the
leading twist contributions to exclusive processes can be combined with
non-perturbative hadron wavefunction information.

The least-complicated exclusive amplitudes
to analyze from first principles in QCD are the space-like
electromagnetic form factors of hadrons. An elastic form factor is
the probability amplitude for a hadron to remain intact
after absorbing momentum $q$ by its local quark current.
If  one uses  light-cone quantization in the $q^+=
q^0+q^z =0$ frame with ${\qp}^2=-q^2=Q^2$,
then vacuum fluctuation contributions
to the $j^+$ current can  be avoided. Nevertheless, the computation of an
elastic form factor  requires knowledge of all of the hadron's light-cone
Fock state wavefunctions.
For example, the helicity-conserving form factor has
the form\refmark\refI\
 $$\eqalign{%
   F(Q^2) &= {\VEV{p+q \vert j^+ \vert p}\over/2p^+} \cr\crr &=
   \sum_{n,\lambda_i} \sum_a e_a \int
   \overline{\prod_i} \  { dx_i\,d^2\kp_i \over  16\pi^3}
   \,\psi_{n}^{(\Lambda)*}(x_i,\lp_i,\lambda_i)
   \,\psi_{n}^{(\Lambda)}(x_i,\kp_i,\lambda_i) . \cr}
 $$ 
The constituents in the initial state have
longitudinal light-cone momentum fractions $x_i=(k^0+k^z)_i/(p^0+p^z),$
relative transverse momentum, $\kp_i,$ and   helicities $\lambda_i.$
Here $e_a$ is the charge of the struck quark, $\Lambda^2\gg\qps$, and the
transverse momenta in the final state are
$$
  \lp_i \equiv \cases{
  \kp_i - x_i \qp + \qp  & \hbox{for the struck quark} \cr
  \kp_i - x_i \qp        & \hbox{for all other partons.}\cr}
$$ 
In principle, one can obtain all of the required Fock State
wavefunctions by
diagonalizing the light-cone QCD Hamiltonian.\refmark\DLCQ\
This has in fact been done for meson and baryon wavefunctions
in the case of QCD in one-space and
one-time dimensions, but the corresponding task appears to be  formidable
for QCD(3+1).

Fortunately, because of asymptotic freedom
and the point-like behavior of quark
and gluon  interactions  at short distances, the computation
of exclusive amplitudes in QCD becomes
much simpler at large momentum transfer. The primary
ingredient in the analysis is factorization: the non-perturbative
dynamics of the bound states can be isolated in terms of
process-independent distribution amplitudes, and the
dynamics of the momentum transfer to the hadrons can be isolated
in terms of perturbatively-calculable hard-scattering quark
and gluon subprocesses. Thus
general properties of exclusive reactions at large momentum
transfer can be derived without explicit knowledge of the
non-perturbative structure of the theory.\refmark\BL\

The most characteristic feature of an
exclusive amplitude in QCD is that it falls
off slowly with momentum transfer, not as an exponential or a
Gaussian, but as an inverse power of $Q=p_T$
which is directly related to the degree of complexity of
the scattering hadrons.  The nominal power-law fall-off\refmark\BF~
${\cal M} \sim
Q^{4-n}$ of an exclusive amplitude at large momentum transfer reflects
the elementary scaling of the lowest-order connected quark and gluon
tree graphs obtained by replacing  each of the external
hadrons by its respective
collinear quarks. Here $n$ is the total number of initial state and
final state lepton, photon, or quark fields entering or leaving the
hard scattering subprocess.   The empirical success of  the
dimensional counting rules for the power-law
fall-off of form factors and general fixed center-of-mass angle
scattering amplitudes gave early and important
evidence for the scale-invariance of quark and gluon interactions
at short distances.

Thus only the valence-quark Fock components of the hadron
wavefunctions contribute to the leading power-law fall-off of
an exclusive amplitude.  In particular, since the internal momentum
transfer at the quark level is required to be large, one can obtain
the basic scaling and helicity structure of the
hadron amplitude by simply iterating the gluon-exchange term in
the effective potential for the light-cone wavefunctions.
The result is that exclusive amplitudes at high momentum transfer $Q^2$
can be written  in a factorized form as a convolution of
process-independent ``distribution amplitudes'' $\phi(x_i,Q)$,
one for each hadron involved in the amplitude, with a hard-scattering
amplitude $T_H$ describing the scattering of the valence
quarks from the initial to final state.\refmark{\refD,\refP}\

The distribution
amplitude is the fundamental gauge invariant  wavefunction
which describes the fractional longitudinal momentum distributions of
the valence quarks in a hadron integrated over transverse momentum
up to the scale $Q$.\refmark\refD\
For example, the pion's electromagnetic form factor
can be written as\refmark{\refD,\refP,\refQ}\
$$
  F_\pi(Q^2) = \int_0^1 dx \int_0^1 dy\,\phi^*_\pi(y,Q)\,T_H(x,y,Q)
  \,\phi_\pi(x,Q)\,\left( 1 + {\cal O}\left({1\over Q}\right)\right) .
  $$
Here $T_H$ is the scattering amplitude obtained when
pions replaced by collinear $q\bar q$ pairs.
This  factorized form is the prototype for the factorization of
general exclusive amplitudes in QCD at high momentum transfer.
All of the non-perturbative dynamics is factorized into the
distribution amplitudes,\refmark\refD~
$\phi_B(x_i,\lambda_i,Q),$ for the baryons with $ x_1+ x_2+ x_3=1,$
and $\phi_M(x_i,\lambda_i,Q),$ for the mesons
with $ x_1+ x_2 =1$ which sum all
internal momentum transfers up to the scale $Q^2.$  On the other
hand, all momentum transfers higher than $Q^2$ appear in $T_H$,
which can be computed perturbatively in powers of
the QCD running coupling constant $\alpha_s(Q^2).$
The distribution amplitudes
are thus the process-independent hadron wavefunctions which interpolate
between the QCD  bound state and their  valence quarks at
transverse separation $b_\perp \simeq 1/Q.$
The pion's distribution amplitude, for example,  is
directly related to its valence light-cone wavefunction:
$$\eqalignno{
\phi_\pi(x,Q) & = \int {d^2\kp \over 16\pi^3}\,\psi^{(Q)}_{q\bar q/\pi}
    (x,\kp)
  &\cr\cr 
 {} & =   P^+_\pi \int {dz^-\over 4\pi}\, e^{i x P^+_\pi z^-/2} \,
  \VEV{0\,\bigg|\,\bar\psi(0)\,{\gamma^+\gamma_5 \over 2\sqrt{2n_c}}
  \,\psi(z)\,\bigg|\,
        \pi}^{(Q)} \bigg|_{\textstyle z^+=\vvec z_\perp=0}.
   &\cr} 
$$
The $\kp$ integration is cut off by the ultraviolet cutoff
$\Lambda=Q$ implicit in the wavefunction; thus
only valence Fock states with invariant mass squared
$\M^2 \leq Q^2$ contribute.

Given the factorized structure of exclusive amplitudes
at large momentum transfer, one can read off a number of general
features of  the PQCD predictions: the dimensional counting
rules,
hadron helicity conservation, and color transparency.\refmark\BL\
QCD
also predicts calculable corrections to the nominal dimensional
counting power-law  behavior
due to the running of the strong coupling constant, higher order
corrections to the hard scattering amplitude, Sudakov effects,
pinch singularities,  as well as the
evolution of the hadron distribution amplitudes, $\phi_H(x_i,Q).$

Evolution equations for the meson and baryon distribution amplitudes
can be derived and employed in analogy to the evolution of
structure functions.\refmark {\BL,\CZ}\
If one can calculate the distribution amplitude
at an initial scale $Q_0$ using QCD sum rules or lattice
gauge theory,\refmark\CZ~ then one can determine $\phi(x_i,Q)$ at higher
momentum scales via evolution equations
in $\log Q^2$ or equivalently, the
operator product expansion.\refmark\refX~
Empirical constraints on the hadron distribution
amplitudes can be obtained from the normalization and scaling of
form factors
at large momentum transfer and  the angular dependence of
two body scattering amplitudes.

Perhaps the most surprising feature of the QCD predictions for exclusive
processes in QCD is ``color transparency",\refmark\CT~
which reflects the fact that only the
small transverse separation $b_\perp
\sim 1/Q$ valence wavefunction can contribute
to exclusive amplitude at large momentum transfer.
Since these color-singlet states have  small color-dipole moments,
they will have small initial and final
state interactions. In particular if the large momentum transfer  occurs
as a quasi-elastic process within a nucleus,
there will be minimal initial state
or final state absorption---in striking contrast
to the standard picture of
strong absorption predicted in  Glauber theory.  A  careful treatment
of color transparency requires consideration
of the expansion time and coherence
length of the small size configurations.\refmark\Miller\

\vglue 0.6cm
\line{\elevenbf 5. A Detailed Example:
Compton Scattering in Perturbative QCD
\hfill}
\vglue 0.4cm

Exclusive reactions involving two real or virtual photons
provide a particularly interesting
testing ground for  QCD because of the relative simplicity
of the couplings of the photons
to the underlying quark currents, and the absence of  significant
initial state interactions---any remnant of
vector-meson dominance contributions
is suppressed at large momentum
transfer, and the photon enters the
amplitude as a direct point-like coupling.

The simplest example of a two-photon exclusive
process is the  $\gamma^*(q) \gamma \to M^0$ process
which is measurable in
tagged $e^+ e^- \to e^+ e^- M^0$ reactions. The
photon to neutral meson transition form factor
$F_{\gamma \to M^0}(Q^2)$ is predicted to fall as
$1/Q^2$---modulo calculable
logarithmic corrections from the evolution
of the meson distribution amplitude.
This QCD prediction reflects
the elementary scaling of the quark propagator at high momentum
transfer, the same scale-free behavior which leads to
Bjorken scaling of the deep inelastic lepton-nucleon
cross sections.  The existing data from TPC/$\gamma\gamma$
are consistent with the predicted scaling
and normalization of
the transition form factors for the $\pi^0, \eta_0,$ and $\eta'.$

The angular distributions for  the
hadron pair production processes $\gamma \gamma \to H \bar H$
are sensitive to the $x_i $ dependence of the hadron distribution
amplitudes.\refmark\BBL\
Lowest order predictions for meson pair production in
two photon collisions using this formalism are given in Refs. \BBL\
and \CZ;
the analysis of the $\gamma \gamma$ to meson pair process
has  been carried out to next-to-leading order in $\alpha_s(Q^2)$
by Nizic.\refmark\Nizic~  The Mark II
and TPC/$\gamma\gamma$  measurements of
$\gamma \gamma \to \pi^+ \pi^-$ and $\gamma \gamma \to
K^+ K^-$ reactions are also consistent with PQCD expectations.
A review of this work is given in Ref. \BREV.

\REF\BLM{%
S. J. Brodsky, G. P. Lepage, and P. B. Mackenzie,
{\it Phys. Rev.} {\bf D28} (1983) 228.}

\REF\CSR{%
S. J. Brodsky, H. J. Lu, SLAC-PUB-6481,  (1994).}

Compton scattering $\gamma p \to \gamma p$
at large momentum transfer and its s-channel crossed reactions
$\gamma \gamma \to \bar p p$ and
$\bar p p \to \gamma \gamma$ are  classic tests of
the perturbative QCD formalism for exclusive reactions. At leading
twist,
each helicity amplitude has the factorized form,\refmark\BL~
$${\cal M}^{\lambda \lambda'}_{h h'}(s,t)=
\sum_{d, i} \int [dx] [dy] \phi_i(x_1,x_2,x_3,\widetilde Q)
T^{(d)}_i(x,h,\lambda; y, h', \lambda';s,t)
\phi_i(y_1, y_2, y_3; \widetilde Q)\ .$$
The index $i$ labels
the three contributing valence Fock amplitudes at
the renormalization scale $\widetilde Q.$
The index $d$ labels the 378 connected Feynman diagrams which
contribute to the eight-point hard scattering amplitude
$q q q \gamma   \to q q q  \gamma$  at the tree level; \ie\  at
order $\alpha \alpha^2_s(\widehat Q).$
The arguments $\widehat Q$ of the QCD running coupling
constant  can be evaluated amplitude by amplitude
using the methods
of Ref. \BLM and \CSR~ as discussed in the Introduction.
The evaluation of the hard scattering amplitudes
$T^{(d)}_i(x,h,\lambda; y, h', \lambda';s,t)$ has now
been done by several groups.\refmark{\FARRAR,\GUNION,\KN,\HYER}\

An important simplification of Compton scattering in PQCD
is the fact that
pinch singularities are readily integrable and do not change the
nominal power-law behavior of the basic amplitudes.\refmark\KN\
Physically,
the pinch singularities correspond to the existence of potentially
on-shell intermediate
states in the hard scattering amplitudes. This
leads to a non-trivial phase structure of the Compton amplitude.
Such phases can in principle be measured by interfering
the virtual Compton process in $e^\pm p \to e^\pm p \gamma$ with
the purely real Bethe-Heitler bremsstrahlung amplitude.\refmark\INT\
A careful analytic treatment of the integration over the on-shell
intermediate states
has been given by Kronfeld and Nizic.\refmark\KN\

The most characteristic feature of the PQCD predictions
is the scaling of the differential Compton cross section at fixed
$t/s$ or $\theta_{CM}$
$$ s^6 {d\sigma \over dt} (\gamma p \to \gamma p) =
F\left(t\over s\right)\ .$$
The power $s^6$ reflects the fact that 8 elementary fields
enter or leave the hard scattering subprocess.\refmark\BF\
The scaling of the existing data\refmark\Shupe~
is remarkably consistent with the PQCD  power-law
prediction, but measurements at higher energies
and momentum transfer are needed to test the predicted logarithmic
corrections to this scaling behavior and determine
the angular distribution of the scaled cross section
over as large a range as possible.

The
predictions for the normalization of the Compton cross section and
the shape of its angular distribution are sensitive to
the shape of the proton distribution amplitude $\phi_p(x_i,Q).$
The forms predicted for the proton distribution amplitude from
QCD sum-rule constraints\refmark\CZ\
by Chernyak, Oglobin, and Zhitnitskii, and  King and Sachrajda,
appear to
give a reasonable representation of the existing data.
A definitive prediction for the normalization of
form factors and other exclusive amplitudes in
perturbative QCD will require not only a careful analysis of
the non-perturbative input for
the distribution amplitudes, but also
a detailed calculation of the crossed-graph
and other irreducible contributions to the hard-scattering QCD kernels.

More recent QCD sum rule analyses of the proton distribution amplitude
are given in Ref. \Stefanis.
These distributions, which predict that approximately
65\%\ of the proton's
momentum is carried by the $u$ quark with helicity parallel
to the proton's helicity also provide empirically consistent
predictions for the normalization of the proton's form factor
and the $J/\psi \to p \bar p$ decay rate.
The crossing behavior from spacelike Compton scattering
to the timelike annihilation
channels will also provide important tests and constraints on the
PQCD formalism and the shape
of the proton distribution amplitudes.
Predictions for the time-like processes have been made by
Farrar {\it et al.},\refmark\FARRAR\ Millers and
Gunion\refmark\GUNION, and Hyer.\refmark\HYER\

The theoretical uncertainties
from finite nucleon mass corrections, the magnitude of the QCD
running coupling constant, and the normalization of the
proton distribution amplitude largely cancel out in the
ratio of Compton to elastic differential cross sections
$$R_{\gamma p /  e^- p }(s, \theta_{cm})=
{{d\sigma( \gamma p \to \gamma p  )\over dt} \bigg/
{d\sigma(  e^-  p \to e^-  p )\over dt}},$$
which is predicted by $QCD$ to be essentially independent of $s$
at large momentum transfer.
If this scaling continues to be confirmed,
then the center-of-mass angular dependence of
$R_{\gamma p /  e^- p }(s, \theta_{cm})$
will be one of the best ways
to determine the shape of $\phi_p(x_i,Q)$.
\eject

\vglue 0.6cm
\line{\elevenbf 6. Lepto-Production of Vector Mesons
as a Test of PQCD \hfill}
\line{\elevenbf \ \ \ \  and Color Transparency
\hfill}
\vglue 0.4cm

The study of real and virtual photoproduction of vector mesons on protons
and nuclei provides an elegant illustration of the emergence
of perturbative QCD features in the large momentum transfer
domain.\refmark{\BFGMS,\SJBexcl,\NNN}\
\pointbegin
At small momentum transfer and high energy where the coherence length
${2 \nu /( {\cal M}^2+ Q^2)}$ is large compared to the target size, the
incident photon is expected to act as a coherent sum of vector mesons
with mass squared ${\cal M}^2 \leq {\cal O} (Q^2).$
This is the
generalized vector meson dominance picture of photon interactions.  In
addition, $s-$channel helicity conservation predicts that the vector
meson will be dominantly produced with transverse polarization equal to
that of the incident photon.
\point
At small momentum transfer where
photon interactions are dominantly hadron-like, the cross section for
vector meson photoproduction on a nucleus should have the same nuclear
properties as meson-nucleon scattering.  Due to the optical theorem, the
forward high energy coherent nuclear amplitude $\gamma^* A \to V^0 A$
must then scale with the nuclear size the same as the total
hadron-nucleus cross section; \ie\  $A^{2/3}.$ The $t-$dependence of
the coherent nuclear cross section is of the form $d\sigma/dt \sim
\exp^{b_A t}$ where $b_A \propto R_A^2$ and $R_A$ is the nuclear size.
Thus the total coherent cross section $\sigma(\gamma^* A \to V^0 A)$ is
predicted to scale with nuclear number as $A^{4/3}/R_A^2 \sim A^{2/3}.$ %
\point
The predictions for $\gamma^* A \to V^0 A'$ are in  striking
contrast to the above results
when $Q^2$ becomes large compared to $\Lambda^2_{QCD}.$
The virtual quark loop connecting the photon to the vector meson is
now  highly
virtual, and only the point-like piece of the photon and the small
transverse size of the valence $q \bar q$ light-cone wavefunction of the
vector meson enter the exclusive amplitude.  Thus at high $Q^2$ the
nuclear absorption in the initial and final state should vanish, and the
nuclear amplitude becomes additive: $M({\gamma^* A \to V^0 A'}) = A^1
M({\gamma^* N \to V^0 N'}).$ The integrated coherent cross section
$\sigma(\gamma^* A \to V^0 A)$ is thus predicted to scale with nuclear
number as $A^{2}/R_A^2 \sim A^{4/3}.$ This contrasting nuclear dependence
of the virtual photoproduction cross section provides a dramatic test of
color transparency.
Preliminary results from E665\refmark\Fang~ for $\rho$
lepto-production at Fermilab appear to confirm these QCD predictions.
\point
Another important prediction of PQCD in the large $Q^2$ domain is
that the vector meson should be produced with zero helicity since it is
formed from a quark and antiquark with equal
and opposite helicities.\refmark\Hel~
The change-over from transverse to longitudinal vector meson polarization
with increasing $Q^2$ also appears to be confirmed by the E665 data.
\point
At large photon virtuality $Q^2$ the photon and vector meson will
act as point-like systems, and thus the $t-$ dependence of the
differential cross section $d\sigma/dt (\gamma^* p \to V^0 p') $ should
only reflect the finite size of the scattered nucleon.
At large $t$ the form factors should reflect the underlying
two-gluon exchange structure of the PQCD Pomeron.
\point
At large
momentum transfer $-t \gg \Lambda^2_{QCD},$ $-u \gg \Lambda^2_{QCD},$
PQCD predicts that the photoproduction cross section has the nominal
fixed CM angle scaling: $d\sigma/dt(\gamma p \to V^0 p') \sim
{f(\theta_{CM})/ s^7}.$ The dominant amplitudes will conserve hadron
helicity: $\lambda_{p'}+\lambda_V = \lambda_p.$
\point
At larger momentum
transfers $-t > R_A^2,$ one can study quasi-elastic lepto-production in
the nucleus; $d\sigma/dt (\gamma^* A \to V^0 N' X)$ where $X$ represents
a sum over excited nuclear states, but without extra particle production.
When $p_T^2 \gg \Lambda^2_{QCD},$ color transparency predicts the absence
of initial or final state absorption of the incident photon and the
outgoing meson and nucleon.  Thus the quasi-elastic cross section should
approach additivity in nuclear number at large momentum transfer.
As I have emphasized in the Introduction, these illuminating studies
and tests of PQCD can be carried out in detail at CEBAF.

\vglue 0.6cm
\vbox{\line{\elevenbf 7. When Do Leading-Twist Predictions
for Exclusive \hfill}
\line{\elevenbf \ \ \ \
Processes Become Applicable?  \hfil}
\vglue 0.4cm}

The factorized predictions for  exclusive
amplitudes are evidently rigorous
predictions  of QCD at large momentum transfer.
However, it is important to understand the kinematic domain
where the leading twist predictions become valid.
The basic scales of QCD are set by the quark masses and the scale
$\Lambda_{QCD}$ which parameterizes the QCD
running coupling constant. Thus
one normally would expect that the leading power-law predictions
should become dominant at momentum transfers exceeding these parameters.
In the case of inclusive reactions, Bjorken scaling is
already apparent at momentum transfers $Q \sim 1$ GeV or less.

In fact, the data for hadron form factors is consistent with the onset
of PQCD scaling at  momentum transfers of a few GeV.
\refmark\Stoler~
has shown that the measurements of the
transition form factors of the proton to the $N(1535)$
and $N(1680)$ resonances are consistent with the predicted
PQCD $Q^{-4}$ scaling to beyond $Q^2 = 20~GeV^2.$
The normalization is also in reasonable
agreement with that predicted from QCD sum rule
constraints on the nucleon distribution amplitudes,
allowing for uncertainties from higher order QCD corrections.
In the case of the proton to $\Delta(1232)$ transition,
the form factor falls faster that $Q^{-4}.$ This anomalous
behavior is, in fact,  predicted by QCD sum rule constraints,
since unlike the proton, the $\Delta$ has a highly symmetric
distribution amplitude which results
in a small net coupling to
the QCD hard scattering amplitude. The observed scaling pattern
of the transition form factors gives strong
support to the QCD sum rule predictions and PQCD factorization.

Isgur and Llewellyn Smith\refmark\ILS~
and  Radyushkin\refmark\RAD~  have raised the
concern that important contributions to exclusive processes
could arise from the endpoint regions
$x_i \to 1;$  such behavior would imply
the breakdown of PQCD factorization.
For example,  the denominator of the hard
scattering amplitudes, {\it e.g.},
$T_H \propto \alpha_s/[(1-x)(1-y)Q^2]$ for the meson form factor
becomes singular in
the endpoint integration region at $x \sim 1$ and $y \sim 1.$
Such endpoint regions are even further
emphasized when one assumes the strongly asymmetric forms
for the hadron distribution amplitudes derived from
QCD sum rules. However,
it is important to note that these endpoint regimes correspond
to scattering processes where one quark carries nearly all
of the proton's momentum and is at a fixed transverse separation
$b_\perp$  from the spectator quarks.

When a quark which is isolated in space receives
a large momentum transfer $x_i Q,$ it will normally
strongly radiate gluons into the final state due to the displacement
of both its initial and
final self-field, which is contrary to the requirements of
exclusive scattering.   For example, in QED the radiation from the
initial and final state charged lines is controlled
by the coherent sum
$\sum_i (\epsilon\cdot p_i/k\cdot p_i) \eta_i q_i$
where $q_i$ and $p_i$ are the charges
four-momenta of the charged lines, $\epsilon$ and $k$ are
polarization and four-momentum of the radiation,
and $\eta_i=\pm 1$ for initial and final
state particles, respectively.
Radiation will occur for any finite momentum transfer scattering
as long as the  photon's
wavelength is less than the size of the initial and final neutral
bound states. The probability amplitude that radiation does not occur is
given by  rapidly falling Sudakov form factor,  as first discussed by
in Refs. \refD\ and \DM.  An elegant and much more complete
discussion has now been given by Botts and Li and Sterman.\refmark\LS\
The radiation from the colored lines in QCD have similar
coherence properties as in QED:\refmark\BG~   because of the
destructive color interference of the radiators,
the momentum of the radiated gluon in a QCD hard scattering process
only ranges from $k$ of
order $1/b_\perp,$ where color screening occurs,
up to  the  momentum transfer  $x_i Q$ of the scattered quarks.
This analysis and unitarity allows one to  compute
the probability that no radiation  occurs
during the hard scattering.\refmark{\LS,\HYER}~ It
is given by a rapidly falling exponentiated Sudakov form factor
$S= S(x_i Q, b_\perp,$ $ \Lambda_{QCD});$ thus at large $Q$ and
fixed impact separation, the Sudakov factor
strongly suppresses the endpoint contribution.
On the other hand, when $b_\perp = {\cal O} (x_i Q)^{-1},$
the Sudakov form
factor is of order 1, and the radiation leads to logarithmic
evolution and contributions of
higher order in $\alpha_s(Q^2),$ the corrections already
contained in the PQCD predictions.\refmark{\refD,\DM,\SM}\
This is the starting point of the detailed
analysis of the suppression of endpoint contributions to
meson and baryon form factors
and its quantitative effect on the PQCD predictions
recently presented by Li and Sterman.\refmark\LS\  This analysis
has now also been applied
to two-photon reactions and the timelike proton form factor by
Hyer.\refmark\HYER\

Thus the leading PQCD contributions to large momentum
transfer exclusive reactions derive from
wavefunction configurations where the
valence quarks are at small transverse separation $ b_\perp = {\cal O}
(1/k_\perp) = {\cal O} (1/Q)$, the regime where there is
no Sudakov suppression. Furthermore, as noted by Li and Sterman,
the hard scattering amplitude  loses its
singular  endpoint structure if one explicitly retains
the valence quark transverse momenta in the denominators.  For
example, in the case of the pion form factor, the hard scattering
amplitude is effectively modified  to   the form
$$T_H \propto {\alpha_s \over
(1-x)(1-y)Q^2 +{ (k_1^\perp+k_2^\perp)^2}}.$$ The Sudakov effect thus
ensures that the denominators are always protected at large momentum
transfers.  In their numerical studies,
Li and Sterman find that  the pion form factor becomes relatively
insensitive to soft gluon exchange at
momentum transfers beyond $20~\Lambda_{QCD}.$
In the case of the proton Dirac form factor,
the corresponding analysis by Li\refmark\LS~ is in good agreement
with experiment at momentum transfers  greater than 3 GeV.
Thus the leading twist QCD predictions  based on
the factorization of long and short distance physics appear to be
self-consistent and valid for momentum transfers
as low as a few GeV, thus
accounting for the empirical success of
quark counting rules in exclusive process phenomenology.
The Sudakov effect suppression also enhances
the QCD ``color transparency" phenomena, since only small color
singlet wavefunction configurations can scatter at large momentum
transfer without radiation.\refmark\CT\

The extension of the leading order PQCD
analysis to higher orders including
Sudakov effects is technically very challenging.  Thus far,
the next-to-leading $\alpha_s(Q^2)$ corrections to the hard scattering
amplitudes $T_H$ have  been computed for only a few exclusive
processes:  the meson form factor,
the photon-to-meson transition form factors,
and $\gamma \gamma$ to meson pairs.
There are many outstanding theoretical
issues which are being resolved, such as how to extend these
calculations to baryon processes, how to set
the renormalization scale in $\alpha_s$,\refmark{\BLM,\CSR}~
how to implement conformal symmetry and
its breaking,\refmark{\refX,\DMueller}~
and how to formulate and solve the
evolution equations for the hadron
distribution amplitudes to next-to-leading
order.

\REF\JiPang{%
C.-R. Ji, A. Pang, and A. Szczepaniak,
North Carolina State University preprint (1994).}

An important question for evaluating exclusive amplitudes
in the transition region
between hard and soft QCD processes is how
to analytically separate perturbative contributions
from contributions intrinsic to the bound-state wavefunction itself.
The physical amplitude of course must be
independent of the choice separation scale $\mu.$
Recently Ji, Pang, and Szczepaniak have observed that
the  natural variable to make this separation
is the light-cone energy or equivalently the
invariant mass of the off-shell partonic system, rather than
gluon virtuality of $T_H.$
The PQCD contributions from the invariant mass regime
$\mu > $ 1 GeV can then account substantially for the empirical
pion form factor at $Q^2 > 1$ GeV$^2.$
One also expects significant contributions from PQCD
from higher order contributions.

It should be emphasized that the measurements
of the pion form factor from electroproduction at large $Q^2$
are quite uncertain
since they requires extrapolation to the pion $t-$ channel pole.
CEBAF measurements can thus contribute significantly to this
fundamental hadronic measure.

\REF\RG{%
E. C. G. St\"uckelberg and A. Peterman, {\it Helv. Phys. Acta}
{\bf 26} (1953) 499. A.~Peterman,
{\it Phys. Rept.} {\bf 53C} (1979) 157.}

\REF\Nizic{%
A similar method is discussed in B. Nizic,
{\it Phys. Rev.} {\bf D35} (1987) 93.}

\REF\LM{%
G. P. Lepage, P. B. Mackenzie,
{\it Phys. Rev.} {\bf D48} (1993) 2250.}

One of the most significant problems in computing the normalization of
perturbative QCD predictions for exclusive processes is the uncertainty
in setting the renormalization scale $\mu$ of the QCD coupling
$\alpha_s(\mu)$ in the hard scattering amplitude $T_H.$ A related problem
is the question of the corresponding scale to use in evaluating the
hadron distribution amplitudes.

Given a renormalization scheme, the QCD Lagrangian ${\cal L}_{QCD}$ is a
function of the bare parameters $\alpha_s(\mu), m_q(\mu),$ etc.  In
principle, the values of the bare parameters can be fixed given a set of
input measurements.  Thus given a finite number of empirical values, all
other QCD observables should be computable order by order in perturbation
theory.  The relation between the input and output observables must be
independent of the choice of the renormalization scale $\mu$ as well as
the choice of intermediate renormalization scheme.  This invariance of
the predictions for observables under changes of the intermediate
renormalization scheme constitutes the generalized renormalization group
invariance of Peterman and St\"uckelberg.\refmark\RG\

Recently, Hung Jung Lu and I\refmark\CSR~
have shown how this problem can be avoided by directly
relating observables through commensurate scale relations.
The conventional ${\overline MS}$ scheme serves as an intermediary
calculational tool, but it can be
systematically eliminated when relating observables.
For example, the entire radiative corrections to the
annihilation cross section  is expressed as the effective charge
$\alpha_R(Q)$ where $Q=\sqrt s$:
$$ R(Q) \equiv 3 \sum_f Q_f^2 \left[ 1+
{\alpha_R(Q) \over \pi} \right] . $$
Similarly, we can define the entire
radiative correction to the Bjorken sum rule as the effective charge
$\alpha_{g_1}(Q)$ where $Q$ is the lepton momentum transfer: $$ \int_0^1
d x \left[
   g_1^{ep}(x,Q^2) - g_1^{en}(x,Q^2) \right]
   \equiv {1\over 6} \left[g_A \over g_V \right]
   \left[ 1- {\alpha_{g_1}(Q) \over \pi} \right] . $$ We now use the
known expressions to three loops in ${\overline MS}$ scheme and choose
the scales $Q^*$ and $Q^{**}$ to re-sum all quark and gluon vacuum
polarization corrections into the running couplings.\refmark\BLM\
The relative scales
insure that each observable pass through the heavy quark thresholds at
their commensurate physical scales.  The final result is remarkably
simple: $${\alpha_{g_1}(Q) \over \pi} = {\alpha_R(Q^*) \over \pi} -
\left( {\alpha_R(Q^{**}) \over \pi} \right)^2 + \left( {\alpha_R(Q^{***})
\over \pi} \right)^3 + \cdots $$ The fundamental test of QCD is then to
verify empirically that the observables track in both normalization and
shape as given by these relations.  The coefficients in the series (aside
for a factor of $C_F,$ which can be absorbed in the definition of
$\alpha_s$) are actually independent of color and are the same in
abelian, non-abelian, and conformal gauge theory.  The non-Abelian
structure of the theory is reflected in the scales $Q^*$ and $Q^{**}.$
The commensurate scale relations thus provide fundamental tests of QCD
which can be made increasingly precise and  independent of any scheme
or other theoretical convention.

In the case of exclusive processes, the coupling  associated
with each virtual gluon exchange carrying momentum transfer
$\ell^\mu_i$ in the hard-scattering subprocess tree
amplitude $T_H$ can be identified with the running coupling
$\alpha_V(\ell^2_i)$ appearing in the heavy quark
potential.\refmark\Nizic~
We can determine the numerical values for
$\alpha_V(Q^2)$ in many ways: directly from
the heavy quarkonium spectrum and
heavy quark lattice gauge theory\refmark\LM~
or from the commensurate scale relations which connect
the $\alpha_V$ scheme to $\alpha_{\bar MS},$  or effective
charges such as $\alpha_R,$  $\alpha_{g_1},$ the Gross-Llewellyn
Smith sum rule, etc. at their appropriate commensurate scales.
Note that higher order corrections  to the hard scattering
amplitude from crossed graph kernels contribute even
if the theory were conformal invariant; \ie\ even of the coupling
did not run.
A related method can be used to choose the separation scale
which controls the evolution of the hadron distribution amplitudes.
By using this procedure, one should be able to substantially reduce the
uncertainty in form factors and other exclusive processes
from renormalization scale  and  scheme ambiguities.

\vglue 0.6cm
\line{\elevenbf 8. Other Applications
of Large Momentum Transfer Exclusive QCD.
 \hfil}
\vglue 0.4cm

The factorization techniques used to derive the leading-twist behavior of
exclusive amplitudes have general
applicability to processes where hadron
wavefunctions have to be evaluated at far off-shell configurations.
In each of these applications, one can  separate the perturbative
quark and gluon dynamics from momentum transfer higher
than a scale $Q$ from the non-perturbative
long-distance physics contained in the distribution amplitudes
$\phi(x_i,Q).$  For example at $x \sim 1$ the struck quark
in deep inelastic lepton-hadron scattering is kinematically far off shell
and space-like. Thus the leading power law fall off in $(1-x)$
is determined by iterating the gluon exchange kernel in
the valence Fock state wavefunction. In this way one derives ``spectator"
counting rules for the nominal power law behavior [e.g. $G_{q/p}(x) \sim
(1-x)^3$] and helicity-retention rules at $x \to 1.$  The resulting
structure functions connect smoothly to the
behavior of large momentum transfer elastic and
inelastic transition form factors at fixed ${\cal M}^2.$  In fact,
when $(1-x) Q^2$ is fixed, the usual evolution of the structure
functions breaks down and there is no increase in the effective
power beyond that given by the spectator
counting rules. Further discussion may be found in Ref. \Nagoya.

Higher-twist corrections to inclusive
reactions are of two types: coherent
corrections which depend on the multiparticle
structure of hadrons, and single particle corrections, such as mass and
condensate insertions, which affect
single quark or single gluon propagators.  Exclusive processes represent
the completely coherent limit of dynamical
higher twist terms in inclusive reactions.   At fixed $(1-x) Q^2,$ the
multi-quark higher twist contributions can
be computed using the exclusive factorization
analysis, and they contribute at the
same order as the leading twist terms.\refmark{\HIGHER,\BHMT}~
Strong higher-twist corrections  are in fact observed in the
angular and $Q^2-$dependence of
Drell-Yan processes and in deep inelastic lepton scattering at  $x \sim
1.$\refmark\conway~

The factorization techniques used to
derive the leading twist contributions to
form factors can also be applied to the exclusive decays of
heavy hadrons when large momentum
transfers are involved.   An interesting example
of this analysis is ``atomic alchemy",\refmark\BGMW~ \ie\
the exclusive decays of muonic atoms to electronic atoms plus neutrinos.
In this case, the calculation requires the  high momentum
tail of the atomic wavefunctions, which in turn
can be obtained via the iteration of the relativistic
atomic bound-state equations. Again one obtains
a factorization theorem for exclusive atomic transitions where the
atomic wavefunction at the origin plays the role of the
distribution amplitude.

\vglue 0.6cm
\line{\elevenbf 9.
Exclusive Nuclear Processes
\hfil}
\vglue 0.4cm

An ultimate goal of QCD phenomenology is to describe the nuclear force
and the structure of nuclei in terms of quark and gluon
degrees of freedom.

One of the most elegant areas of application of QCD to nuclear
physics is the domain of large momentum transfer exclusive nuclear
processes.
Rigorous results have been given by Lepage, Ji and
myself$\,$\refmark\blj~ for the asymptotic
properties of the deuteron form factor at large momentum transfer.
In the asymptotic $Q^2\rarrow \infty$ limit the deuteron distribution
amplitude, which controls large momentum transfer deuteron reactions,
becomes fully symmetric among the five possible color-singlet
combinations of the six quarks.
One can also study the evolution of the ``hidden color'' components
(orthogonal to the $np$ and $\Delta\Delta$ degrees of freedom)
from intermediate to large momentum transfer scales; the results also
give constraints on the nature of the nuclear force at short
distances in QCD.
The existence of hidden color degrees of freedom further illustrates
the complexity of nuclear systems in QCD.
It is conceivable that six-quark $d^\ast$ resonances corresponds to
these new degrees of freedom may be found by careful searches of
the $\gamma^\ast d\rarrow \gamma d$ and $\gamma^\ast d\rarrow
\pi d$ channels.

The basic scaling law for the helicity-conserving deuteron
form factor is
$F_d(Q^2) \sim 1/Q^{10}$ which comes from simple quark counting
rules, as well as perturbative QCD. One cannot expect
this asymptotic prediction to become accurate until
very large $Q^2$ is reached since the momentum transfer
has to be shared by at least six constituents.
However, one can identify the QCD physics due to the compositeness
of the nucleus, with respect to its nucleon
degrees of freedom by using the reduced amplitude formalism.\refmark\BC~
For example, consider the
deuteron form factor in QCD. By definition this quantity is the
probability amplitude for the deuteron to scatter from
$p$ to $p+q$ but remain intact.%
\Picture\figred\width=\hsize
\caption{\narrower\singlespace
Application of the reduced amplitude formalism to the
deuteron form factor at large momentum transfer.
}
\savepicture\figredpic
\Picture\figyp\width=\hsize
\caption{\narrower\singlespace
Scaling of the deuteron reduced form factor.
The data are summarized in Ref.~\Aan.
}
\savepicture\figyppic
Note that for vanishing nuclear binding
energy $\epsilon_d \rarrow 0$, the deuteron can be regarded as two
nucleons sharing the deuteron four-momentum (see Fig. \figred).
The momentum $\ell$ is limited by the binding and can thus be neglected.
To first approximation
the proton and neutron share the deuteron's momentum equally.
Since the deuteron form factor contains the probability amplitudes
for the proton and neutron to scatter from $p/2$ to $p/2+q/2$;
it is natural to define the reduced
deuteron form factor\refmark{\BC,\Aan,\Aam}\
$$ f_d(Q^2) \equiv {F_d(Q^2)\over F_{1N}
\left(Q^2\over 4\right)\, F_{1N}\,\left(Q^2\over 4\right)}.$$
The effect of nucleon compositeness is removed
from the reduced form factor.   QCD then predicts the scaling
$$ f_d(Q^2) \sim {1\over Q^2} $$
\ie\ the same scaling law as a meson form factor.
Diagrammatically, the extra power of $1/Q^2$ comes from the
propagator of the struck quark line, the one propagator not contained
in the nucleon form factors. Because of hadron helicity conservation,
the prediction is
for the leading helicity-conserving deuteron form factor
$(\lambda=\lambda'= 0.)$  As shown in Fig. \figyp, this scaling
is consistent with experiment for $Q= p_T \gsim$ 1 GeV.

\midinsert
\vskip 2.5in
\figredpic\endinsert

\Picture\figyo\width=\hsize
\caption{\narrower\singlespace
Construction of the reduced nuclear amplitude for two-body
inelastic deuteron reactions.\refmark\Aan\
}
\savepicture\figyopic

The distinction between the QCD and other treatments of
nuclear amplitudes is particularly clear
in the reaction $\gamma d \rightarrow n p$; \ie\
photo-disintegration of the deuteron at fixed center of mass angle.
Using dimensional counting, the leading power-law prediction
from QCD is simply ${d\sigma\over dt}(\gamma d \rightarrow n p) \sim
 F(\theta_{\rm cm})/s^{11}$.
Again we note that the
virtual momenta are partitioned among many quarks and gluons, so that
finite mass corrections will be significant at low to medium energies.
Nevertheless, one can test the basic QCD dynamics in these
reactions taking into account much of the finite-mass,
higher-twist corrections
by using the ``reduced amplitude'' formalism.\refmark{\Aan,\Aam}\
Thus the photo-disintegration amplitude
contains the probability amplitude (\ie\ nucleon form factors)
for the proton and neutron to each remain intact after
absorbing momentum transfers $p_p-1/2 p_d$
and $p_n-1/2 p_d,$ respectively (see Fig. \figyo).
After the form factors
are removed, the remaining ``reduced" amplitude should scale
as $F(\theta_{\rm cm})/p_T$. The single inverse power of
transverse momentum $p_T$ is the slowest conceivable in any theory,
but it is the unique power predicted by PQCD.

\midinsert
\vskip 3.5in
\figyppic\endinsert

\midinsert
\vskip 2in
\figyopic\endinsert

\Picture\figyq\width=\hsize
\caption{\narrower\singlespace
Comparison of deuteron photodisintegration data with
the scaling prediction which requires $f^2(\theta_{cm})$ to
be at most logarithmically dependent
on energy at large momentum transfer.
The data in (a) are from the recent experiment of Ref.~\Holt.
The nuclear physics prediction shown in (a) is from Ref.~\Lee.
The data in (b) are from Ref.~\Myers.
}
\savepicture\figyqpic

The prediction that $f(\theta_{\rm cm})$ is
energy dependent at high-momentum transfer is compared with experiment
in Fig. \figyq. It is particularly striking to see the
QCD prediction verified at incident photon lab energies as low
as 1 GeV. A comparison with a standard nuclear physics model
with exchange currents is also
shown for comparison as the solid
curve in Fig. \figyq (a).
The fact that
this prediction falls less fast than the data suggests that meson and
nucleon compositeness are not taken to into account correctly.
An extension of these data to other angles and higher energy
would clearly be very valuable.

\midinsert
\vskip 3.75in
\figyqpic\endinsert

The derivation of  the evolution equation for the deuteron and other
multi-quark states is given in Refs. \blj\  and \Aam.\
In the case of the
deuteron, the evolution equation couples five different color
singlet states composed of the six quarks.  The leading
anomalous dimension for the deuteron distribution amplitude and
the helicity-conserving deuteron form factor at asymptotic $Q^2$
is given in Ref. \blj.

There are a number of related
tests of QCD and reduced amplitudes which require $\bar p$
beams\refmark\Aam\  such as
$\bar pd \rarrow \gamma n$ and $\bar p d\rarrow \pi  p$
in the fixed $\theta_{\rm cm}$ region.
These reactions are particularly interesting tests of
QCD in nuclei.
Dimensional counting rules
predict the asymptotic behavior ${d\sigma\over dt}\
(\bar pd\rarrow \pi  p)\sim {1\over (p_T^2)^{12}} \ f(\theta_{\rm cm})$
since there are 14 initial and final quanta
involved.  Again one notes that
the $\bar p d \rarrow \pi  p$ amplitude  contains a factor
representing the
probability amplitude (\ie\ form factor) for the proton to remain
intact after absorbing momentum transfer squared $\widehat t =
(p-1/2 p_d)^2$ and the $\bar NN$ time-like form factor at
$\widehat s = (\bar p + 1/2 p_d)^2$.
Thus ${\cal M}_{\bar pd\rarrow \pi  p}\sim F_{1N}(\widehat t)\
F_{1N}(\widehat s)\, {\cal M}_r,$
where
${\cal M}_r$ has the same QCD scaling
properties as quark meson scattering.
One thus predicts $$
{{d\sigma\over d\Omega}\ (\bar pd\rarrow \pi  p)\over
F^2_{1N}(\widehat t)\, F^2_{1N}(\widehat s)} \sim {f(\Omega)\over
p^2_T} \ .
$$
The reduced amplitude scaling for
$\gamma d \rarrow pn$ at large angles and $p_T \gsim$ 1~GeV
(see Fig. \figyq).   One thus expects similar precocious scaling
behavior to hold for $\bar p d \rarrow \pi   p$
and other $\bar pd $ exclusive reduced amplitudes.
An analysis by Kondratyuk and Sapozhnikov\refmark\KOD\  shows
that standard  nuclear physics
wavefunctions and interactions cannot explain the magnitude of the
data for two-body anti-proton annihilation reactions such as
$\bar p d \rarrow \pi   p$.

\vglue 0.6cm
\line{\elevenbf 10.
Outstanding Phenomenological Issues in Exclusive Processes.
\hfil}
\vglue 0.4cm

Although most large momentum transfer exclusive reactions appears
to be empirically consistent with
perturbative QCD expectations, there are a
number of glaring exceptions where theory and experiment diverge.
If one accepts that the underlying formalism for the
leading twist behavior of exclusive reactions is reliable, then
these exceptions provide important insights into new  physical mechanisms
within QCD.

{\sl What accounts for the structure in
the spin correlations in $\it{pp}$ elastic
scattering at large momentum transfer?}\
Measurements\refmark\Krisch~ of large angle
$p p$ elastic scattering at Argonne
and Brookhaven show a dramatic spin-spin correlation
$A_{NN}$ which reaches $
\sim 0.6 $ at $\sqrt s \sim 5$ GeV: \ie~ the spin-analyzed
cross section is four times larger
if the protons scatter with their spins parallel
and normal to the scattering
plane compared to antiparallel. The explanation
for this phenomena is far from
settled. The most popular explanations\refmark\pinch~ are based on the
interference of Landshoff pinch singularities\refmark\Lands~
with  the quark interchange
amplitude, but there is no understanding why the
Landshoff contribution would
itself have a large $A_{NN}$\refmark\Lip~
or sufficient normalization\refmark\Sivers~
to explain this
phenomena. Guy de~Teramond and I have proposed\refmark\BDeT\
that the large spin correlations reflects  inelastic channels
corresponding to the production of
charm at threshold. This effect
leads to enhancement in the $J=L=S=1$ $pp \to pp$ partial
wave which implies
a large value of $A_{NN}$ at the energies sufficient to produce open
charm. This explanation would  be confirmed by the observation
of a sizeable charm production  in $p p$ collisions at a
rate of order of 1 microbarn.
A similar enhancement of $A_{NN}$ is
seen at the open strangeness threshold regime.
and is consistent with the 1 millibarn cross section
observed for the production of strange hadrons just above threshold.
The heavy quark explanation has received some support from the
work of Luke, Savage, and Manohar,\refmark\LMS~ who have shown that the
interactions of $c \bar c$ systems
at low relative velocity with hadrons is
enhanced due to the QCD scale anomaly;
in fact, the scalar exchange interaction
is predicted to be strong enough to bind charmonium to heavy
nuclei.\refmark\BDeTS

{\sl Why does QCD color transparency appear to break down in quasielastic
$\it{pp}$ scattering.} \
The Brook\-haven measurements\refmark\Hepp~
of  the transparency ratio for large
angle quasi-elastic $p p$ scattering increases with momentum transfer, as
predicted by PQCD, but the ratio then
appears to revert to normal absorption at
$\sqrt s \sim 5$ GeV. This suggests that
whatever is causing the structure in
$A_{NN}$ at the same energies and angles
involves large transverse sizes and is
far from perturbative in origin. The charm threshold
effect is a candidate for this type of explanation.

The preliminary results for the SLAC color
transparency experiment NE-18\refmark\NE~ indicate that
color transparency in quasi-elastic $e
p$ scattering  is not a strong effect up
to the  accessible momentum transfers.
Higher momentum transfers exceeding 5 GeV
are needed for a decisive test. A
sensitive test of color transparency is provided by measuring the
sign of the derivative of the transparency
ratio $ {d\sigma dQ^2 (e A \to e' p
(A-1)) \over Z d\sigma dQ^2 (e p \to e' p)}\ .
$ Perturbative QCD predicts a positive slope, whereas
conventional Glauber theory
predicts a negative derivative in the low $Q^2$ domain.

{\sl Why does the $J/\psi$ decay copiously to $\rho \pi$?}\
According to the principle of hadron
helicity conservation\refmark\Hel~  in exclusive decays,
the $J/\psi$  produced with $J_z= \pm 1$
in $e^+ e^-$ annihilation should not
decay to vector plus pseudoscalar meson pairs.
In fact, this is true for the $\psi'$ and other S-state
charmonium states, but in the case of the
$J/\psi,$ the $\rho\pi$ and $K K^*$ psuedoscalar-vector meson
channels are actually the dominant two-body hadronic
decays. A possible explanation is that the $J/\psi$
mixes with a nearby gluonic or hybrid $J=1$ state ${\cal O}$
that favors vector plus pseudoscalar
meson pair decay.\refmark\Tuan~
One can search for the ${\cal O}$ by looking
for a $\rho \pi$  mass peak near the $J/\psi$
in the decay $\psi' \to \pi \pi {\cal O} \to \pi \pi \rho \pi.$

{\sl Why do effective Reggeon trajectories flatten to values below
$\alpha_R(t) = 0$ at large momentum transfer?}\
A fundamental prediction of perturbative QCD
is that the Reggeon trajectories $\alpha_\rho(t)$ and $\alpha_{A_2}(t)$
governing charge exchange reactions at high energies $s \gg -t$
monotonically approach zero
at large spacelike momentum transfer.\refmark\Kirschner~
More generally,
the leading Reggeon in an exclusive process will reflect the minimal
particle number exchange quantum numbers:
two gluons in the case of the Pomeron,
three gluons in the case of the Odderon,
and quark plus anti-quark in the case of
meson exchange trajectories. Because of asymptotic freedom the leading
trajectory at large momentum transfer is thus simply $j_1+j_2-1$
with corrections of order $\sqrt \alpha_s(-t).$
The asymptotic prediction $\lim_{-t \to \infty}
\alpha_R(t) =0$ reflects the fact that
a weakly interacting quark-antiquark pair
is exchanged in the $t-$channel.\refmark\Kirschner~
Thus one expects that the effective
$\rho$ Reggeon should asymptote at
$\alpha_\rho(t) \to 0$ at large $-t.$ However, measurements
of the inclusive processes $\pi^- p \to \pi^0 X$
at $s \simeq 300$ GeV$^2$ and $ 8 > -t > 2 $GeV$^2$
indicate that the effective non-singlet $\rho$ trajectory
becomes negative at large $-t.$\refmark\Savit~ Thorn, Tang and I have
recently shown that the hard QCD part of the trajectory is weakly coupled
and that its contribution may well be hidden until much higher
energy.\refmark\Thorn~  Quark interchange\refmark\BBG~
may thus be the dominant
subprocess at presently accessible kinematic ranges.
We also show that Reggeon contributions to
exclusive and semi-inclusive mesonic exchange hadron
reactions can be systematically studied
in perturbative QCD.

{\sl Why is quark interchange the dominant
mechanism for large-angle hadron-hadron scattering?}\
The comprehensive measurements at BNL\refmark\Carroll~ of the relative
normalization and angular dependence of a large set of exclusive hadron
scattering channels strongly suggests that
the dominant mechanism for scattering
hadrons at large momentum transfer is quark
interchange.\refmark\BBG~ For example,
if gluon exchange were the dominant mechanism,
then the differential cross
sections for  $K^+p \to K^+ p $ and $K^- p
\to K^- p$ at large $p_T$ would be roughly equal in
magnitude and angular shape. In fact they
have grossly different magnitudes and shapes. The
$K^+ p \to K^+ p$  cross section  has the
approximate form predicted by the exchange of their common $u$ quark.
A possible explanation of this fact is that
quark interchange involves the least
number of large momentum exchanges within the
hadron scattering amplitude.

The short-distance structure of hadrons, hadron dynamics,
and hadronization is
thus one of the frontier areas of study in testing
quantum chromodynamics. Electroproduction at
CEBAF will play an essential role in resolving this
fundamental area of physics.

\vglue 0.6cm
\line{\elevenbf  Acknowledgements \hfil}
\vglue 0.4cm
I wish to thank Carl Carlson and Paul Stoler and the other members of the
organizing committee for organizing an outstanding meeting in
Elba.  I also thank Tom Hyer for helpful conversations.
This work was supported by the
U.S. Department of Energy under Contract No. DE-AC03-76SF00515.
\refout

\vfil\supereject
\bye